\documentclass[preprint,11pt]{aastex}

\shortauthors{Winn et al.~2007}
\shorttitle{System Parameters of HD~189733}

\begin{document}

% ------------------------------------------------------------------------
% New commands
%
\def\ltsima{$\; \buildrel < \over \sim \;$}
\def\lsim{\lower.5ex\hbox{\ltsima}}
\def\gtsima{$\; \buildrel > \over \sim \;$}
\def\gsim{\lower.5ex\hbox{\gtsima}}
\def\lam{\lambda=-1\fdg4 \pm 1\fdg1}
% -------------------------------------------------------------------------
%

\bibliographystyle{apj}

\title{
The Transit Light Curve Project.\\
V.~System Parameters and Stellar Rotation Period of HD~189733
}

\author{
Joshua N.\ Winn\altaffilmark{1},
Matthew J. Holman\altaffilmark{2},
Gregory W.\ Henry\altaffilmark{3},
Anna Roussanova\altaffilmark{1},\\
Keigo Enya\altaffilmark{4},
Yuzuru Yoshii\altaffilmark{5,6},
Avi Shporer\altaffilmark{7},
Tsevi Mazeh\altaffilmark{7},\\
John Asher Johnson\altaffilmark{8},
Norio Narita\altaffilmark{9},
Yasushi Suto\altaffilmark{9}
}

\altaffiltext{1}{Department of Physics, and Kavli Institute for
  Astrophysics and Space Research, Massachusetts Institute of
  Technology, Cambridge, MA 02139, USA}

\altaffiltext{2}{Harvard-Smithsonian Center for Astrophysics, 60
  Garden Street, Cambridge, MA 02138, USA}

\altaffiltext{3}{Center of Excellence in Information Systems,
  Tennessee State University, 3500 John A.\ Merritt Blvd., Box 9501,
  Nashville, TN 37209, USA}

\altaffiltext{4}{Institute of Space and Astronautical Science, Japan
  Aerospace Exploration Agency, 3-1-1, Yoshinodai, Sagamihara,
  Kanagawa, 229-8510, Japan}

\altaffiltext{5}{Institute of Astronomy, School of Science, University
  of Tokyo, 2-21-1 Osawa, Mitaka, Tokyo 181-0015, Japan}

\altaffiltext{6}{Research Center for the Early Universe, School of
  Science, University of Tokyo, 7-3-1 Hongo, Bunkyo-ku, Tokyo
  113-0033, Japan}

\altaffiltext{7}{Wise Observatory, Raymond and Beverly Sackler Faculty
  of Exact Sciences, Tel Aviv University, Tel Aviv 69978, Israel}

\altaffiltext{8}{Department of Astronomy, University of California,
  Mail Code 3411, Berkeley, CA 94720, USA}

\altaffiltext{9}{Department of Physics, The University of Tokyo,
  Tokyo 113-0033, Japan}

\begin{abstract}

  We present photometry of HD~189733 during eight transits of its
  close-in giant planet, and out-of-transit photometry spanning two
  years. Using the transit photometry, we determine the stellar and
  planetary radii and the photometric ephemeris. Outside of transits,
  there are quasiperiodic flux variations with a 13.4~day period that
  we attribute to stellar rotation. In combination with previous
  results, we derive upper limits on the orbital eccentricity, and on
  the true angle between the stellar rotation axis and planetary orbit
  (as opposed to the angle between the projections of those axes on
  the sky).

\end{abstract}


\section{Introduction}

For the same reason that eclipsing binary stars are important in
stellar astrophysics, transiting planets play an outsized role in
exoplanetary science. This can be appreciated by comparing the
transiting exoplanet HD~189733b (Bouchy et al.~2005), the subject of
this paper, to the arbitrarily chosen non-transiting planet HD~187123b
(Butler et al.~1998). Both planets are ``hot Jupiters'' detected by
the Doppler method. All that is known about HD~187123b is its orbital
period ($P=3.097$~days) and minimum mass ($M_p\sin i = 0.52$~$M_{\rm
  Jup}$), despite 6~years having elapsed since its discovery. In
contrast, for HD~189733b, discovered only 1.5~yr ago, transit
photometry has revealed the planet's radius ($1.15$~$R_{\rm Jup}$;
Bouchy et al.~2005, Bakos et al.~2006a) and removed the $\sin i$
ambiguity in the planet's mass ($M_p = 1.13$~$M_{\rm Jup}$). Infrared
photometry during a secondary eclipse has led to a determination of
the planet's 16~$\mu$m brightness temperature ($T_p=1300~K$; Deming et
al.~2006). Most recently, spectroscopic observations during a transit
have shown that the angle on the sky between the orbit normal and the
stellar rotation axis is within a few degrees of zero (Winn et
al.~2006).

These measurements are essential for a complete understanding of the
atmospheres and interiors of hot Jupiters, as well as their formation
and migration mechanisms. An even richer set of measurements can be
expected in the future, from investigators pursuing transmission
spectroscopy, reflected-light observations, and other transit-related
investigations (as reviewed recently by Charbonneau et al.~2006a). One
goal of the Transit Light Curve (TLC) Project is to support these
efforts by refining the estimates of the planetary, stellar, and
orbital parameters, through high-accuracy, high-cadence photometry of
exoplanetary transits. We also seek to measure or bound any variations
in the transit times and light-curve shapes that would be caused by
the influence of additional bodies in the system (Miralda-Escud\'{e}
2002; Agol et al.~2005; Holman \& Murray 2005). Along the way, we are
exploring different techniques for photometry and parameter
determination. Previous papers in this series have reported results
for the exoplanets XO-1b (Holman et al.~2006a), OGLE-TR-111b (Winn et
al.~2007), TrES-1 (Winn, Holman, \& Roussanova 2006), and OGLE-TR-10b
(Holman et al.~2006b).

This paper presents our results for HD~189733b, along with
out-of-transit photometry spanning two years. The reason for gathering
out-of-transit photometry was to attempt to measure the stellar
rotation period. The parent star is relatively active, with a
chromospheric activity index $S=0.525$ (Wright et al.~2004) and
$\log{R'_{HK}} = -4.4$, raising the possibility of measuring the
rotation period through starspot-induced quasiperiodic flux
variations. As we will explain, the measurements of both the rotation
period and the Rossiter-McLaughlin effect permit the determination of
the true angle between the orbit normal and stellar rotation axis (as
opposed to the angle between the projections of those vectors on the
sky, which was measured by Winn et al.~2006).

This paper is organized as follows. In \S~2, we present photometry of
8 different transits, along with nightly photometry over two
consecutive observing seasons. In particular, \S~2.2 presents the
out-of-transit photometry and the estimation of the stellar rotation
period. In \S~3, we describe the parameteric model that was fitted to
the data, and in \S~4 we present the results for the planetary,
stellar, and orbital parameters, as well as the new transit
ephemerides, a limit on the orbital eccentricity, and the
three-dimensional spin-orbit alignment. The last section is a summary.

\section{Observations and Data Reduction}

Our observations took place in 2005 and 2006 using telescopes at 4
different observatories.  Each of the sections below presents the
photometry from a given observatory.  All together, we observed 8
different transits. Table~1 gives a summary of the characteristics of
the data from each transit, and Table~2 gives the final photometry. A
telescope at Fairborn Observatory (\S~2.2) was also used to monitor
the out-of-transit flux of HD~189733 over the two observing seasons.

\subsection{Fred Lawrence Whipple Observatory}

We observed the transits of UT~2006~Jul~21 and Sep~10 with the 1.2m
telescope at the Fred L.\ Whipple Observatory on Mt.\ Hopkins,
Arizona.\footnote{The data from the first of these two transits have
  already been presented by Winn et al.~(2006).} We used KeplerCam,
which has one 4096$^2$ Fairchild 486 back-illuminated CCD, with a
$23\farcm1 \times 23\farcm 1$ field of view. For our observations we
used $2\times 2$ binning, which gives a scale of $0\farcs 68$ per
binned pixel, a readout and setup time of 11~s, and a typical readout
noise of 7 e$^-$ per binned pixel. We observed both transits through
the SDSS $z$~band filter, in order to minimize the effect of
color-dependent atmospheric extinction on the relative photometry and
to minimize the effect of limb-darkening on the transit light
curve. We deliberately defocused the telescope such that the
full-width at half-maximum (FWHM) of stellar images was about 7~binned
pixels ($4\farcs8$) in order to permit a consistent exposure time of
5~s. We used automatic guiding to keep the image registration as
constant as possible. We also obtained dome flat exposures and
zero-second (bias) exposures at the beginning and the end of each
night.

On UT~2006~Jul~21, the sky conditions began partly cloudy but
gradually improved as the night went on. We observed the target star
as it rose from an airmass of 1.19 to 1.01 and then descended to an
airmass of 1.23.  There was a 15-minute interruption after second
contact, due to clouds.  On UT~2006~Sep~10, the sky conditions were
mainly clear (but not all-sky photometric). We followed the target
star from an airmass of 1.02 to 2.5, although we discarded the data
taken at airmass~$> 1.9$ because of their much poorer quality. There
was a 15-minute interruption prior to third contact, due to a computer
crash, which also caused a change in image registration and focus.

We used standard IRAF\footnote{ The Image Reduction and Analysis
  Facility (IRAF) is distributed by the National Optical Astronomy
  Observatories, which are operated by the Association of Universities
  for Research in Astronomy, Inc., under cooperative agreement with
  the National Science Foundation.  } procedures for the overscan
correction, trimming, bias subtraction, and flat-field division. We
performed aperture photometry of HD~189733 and 14 nearby and
necessarily fainter stars. The light curve of each comparison star was
normalized to have unit median, and the mean of these normalized light
curves was taken to be the comparison signal. The light curve of
HD~189733 was divided by the comparison signal, and corrected for
residual systematic effects by dividing out a linear function of time.
The zero point and slope of the linear function were determined as
part of the model-fitting procedure, as explained in \S~4. Figure~1
shows the final light curves, along with a time-averaged composite
light curve created from the two data sets.

\begin{figure}[p]
\epsscale{1.0}
\plotone{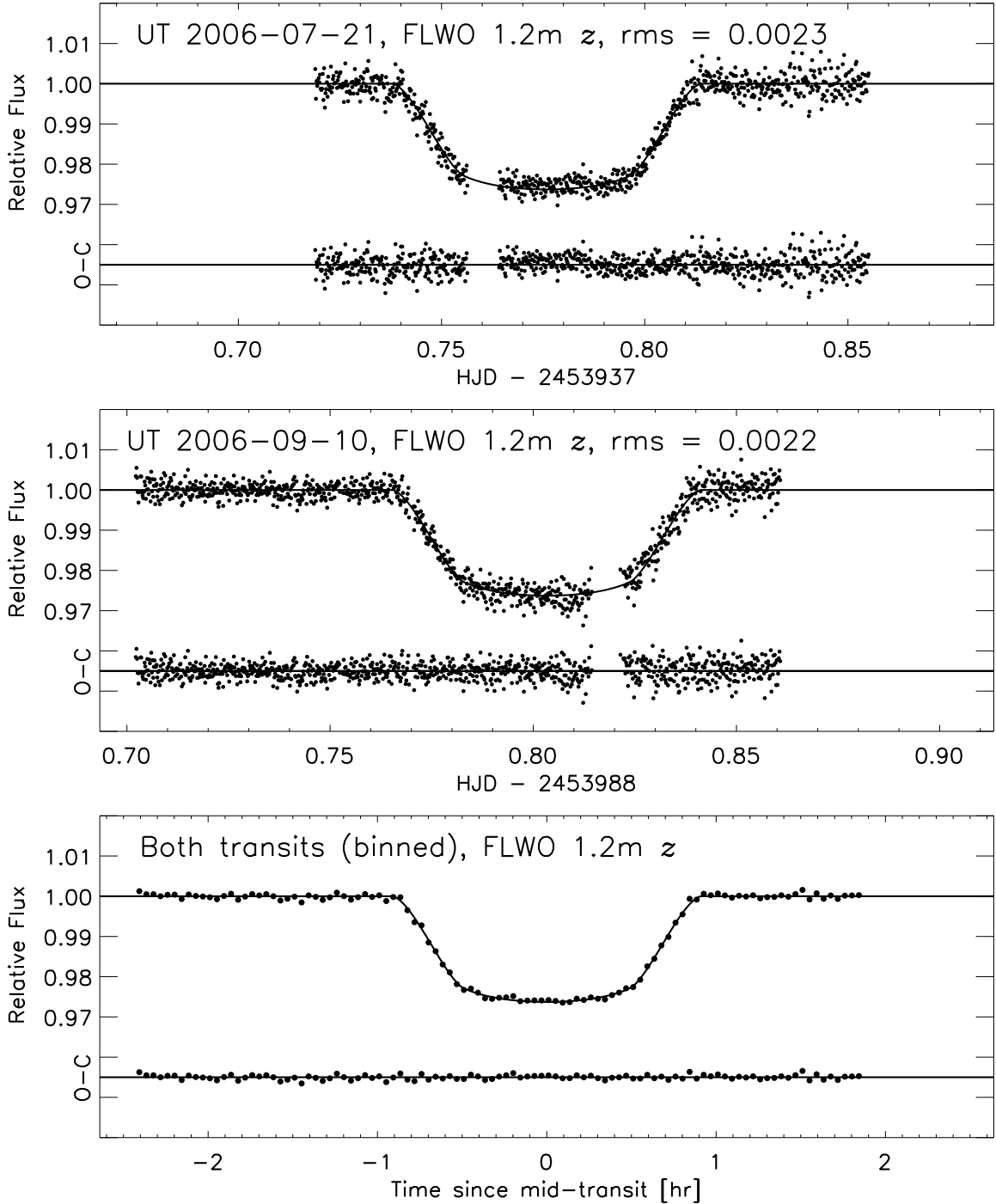}
\vskip 0.2in
\caption{
Photometry of HD~189733 in the $z$ band, using
the FLWO 1.2m telescope and Keplercam.  These data were used to
estimate the planetary, stellar, and orbital parameters (see \S~3).
The bottom panel is a composite light curve created from both data
sets, after time-shifting the earlier transit and averaging into
1~minute bins.  The residuals (observed$-$calculated)
are plotted beneath the data.
\label{fig:1}}
\end{figure}

\subsection{Fairborn Observatory}

We used the T10 0.8m automated photometric telescope (APT) at Fairborn
Observatory, in Arizona, to observe four complete transits of
HD~189733b and to monitor the out-of-transit stellar flux.  The T10
APT is equipped with a two-channel precision photometer employing two
EMI 9124QB bi-alkali photomultiplier tubes to make simultaneous
measurements in the Str\"omgren $b$ and $y$ passbands.  The APT
measures the difference in brightness between a program star and a
nearby constant comparison star (or stars) with a typical precision of
0.0015~mag for bright stars ($V < 8.0$). For the HD~189733 transits,
we used the comparison star HD~189410 ($V$ = 5.68, $B-V$ = 0.34, F0).
The differential magnitudes were reduced with nightly extinction
coefficients and transformed to the Str\"omgren system with yearly
mean transformation coefficients.  To improve our photometric
precision, we combined the separate $b$ and $y$ differential
magnitudes into a single $(b+y)/2$ passband.  For additional
information on the telescopes, photometers, observing procedures, and
data reduction techniques, see Henry~(1999) and Eaton, Henry, \& Fekel
(2003).

We observed the transits of UT~2005~Nov~28, 2006~May~2, 2006~May~22,
and 2006~Jun~11.  On each of those nights, the differential magnitude
of HD~189733 was recorded for 3-5 hours bracketing the expected time
of mid-transit.  The transit light curves are shown in the top 4
panels of Fig.~2.  Like the FLWO data, these data have been corrected
by a linear function of time that was determined as part of the
fitting procedure (see \S~4).

In addition, we also measured the out-of-transit flux with the APT on
93 different nights spanning two observing seasons between
2005~October and 2006~July.  Three comparison stars were observed on
each night.  The out-of-transit flux measurements are shown in the top
two panels of Fig.~3.  The flux varied erratically during the 2005
observing season and at the beginning of the 2006 season.  However, a
quasiperiodic signal became evident near the end of our 2006 season
observations with a peak-to-peak amplitude of 1.3\% and a period of
about 13~days, although only 2.5~cycles were observed.

This type of photometric behavior---erratic and occasionally
quasiperiodic variation---is common among chromospherically active
stars, especially stars with low or intermediate levels of activity
(see, e.g., Henry, Fekel, \& Hall 1995). The photometric variability
in these stars arises from photospheric starspots and plages that are
carried into and out of view by the stellar rotation.  In the case of
HD~189733, the spots cover only $\sim$1\% of the stellar surface at
any moment, and their distribution on the star changes significantly
on the rotational timescale.  A periodogram of the quasiperiodic
portion of the light curve (indicated by the solid dots in the middle
panel of Fig.~3) shows a strong peak at 13.4~days, which we take to be
the stellar rotation period. We estimated the uncertainty by
recalculating the light curve using each of the three different
comparison stars; the standard deviation of the results was 0.4~days.
Thus, in what follows, we adopt the value $P_{\rm rot} = 13.4 \pm
0.4$~days. Since this result is only based on 2.5~cycles, continued
observations are warranted to check our estimate. There may be
additional errors because of differential rotation and the variations
in spot positions and intensities on the rotational timescale. The
bottom panel of Fig.~3 shows the flux as a function of rotational
phase during the epoch of quasiperiodic variation.  A periodogram of
the entire data set also shows a peak at 13.4~days, along with a peak
at 6.7~days, presumably from a time period when star spots occurred on
both sides of the star.

\begin{figure}[p]
\epsscale{0.5}
\plotone{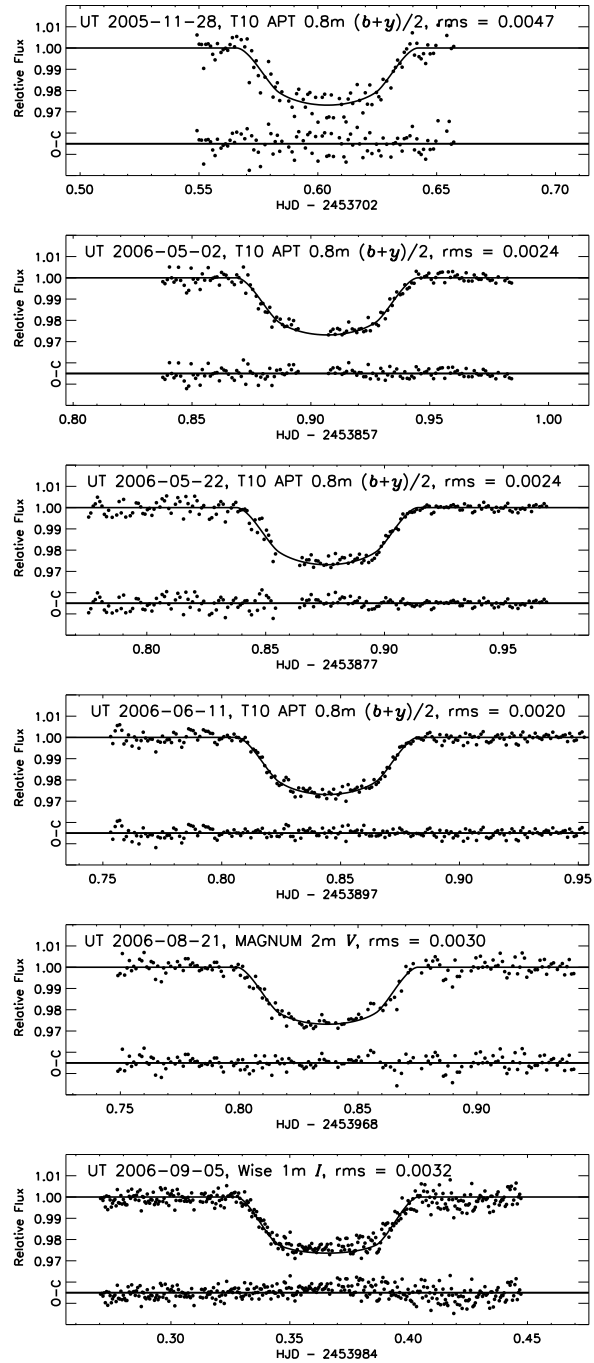}
\caption{
Relative photometry of HD~189733 during 6 different transits. The date,
telescope, and filter are identified on each panel. These data were used only for
measurements of the times of transit (see \S~3). The model, shown as
the solid line, is based on the model derived from the FLWO~$z$-band
data after changing the limb-darkening parameters appropriately. In
all cases, the residuals (observed$-$calculated) are plotted beneath
the data.
\label{fig:2}}
\end{figure}

\begin{figure}[p]
\epsscale{1.0}
\plotone{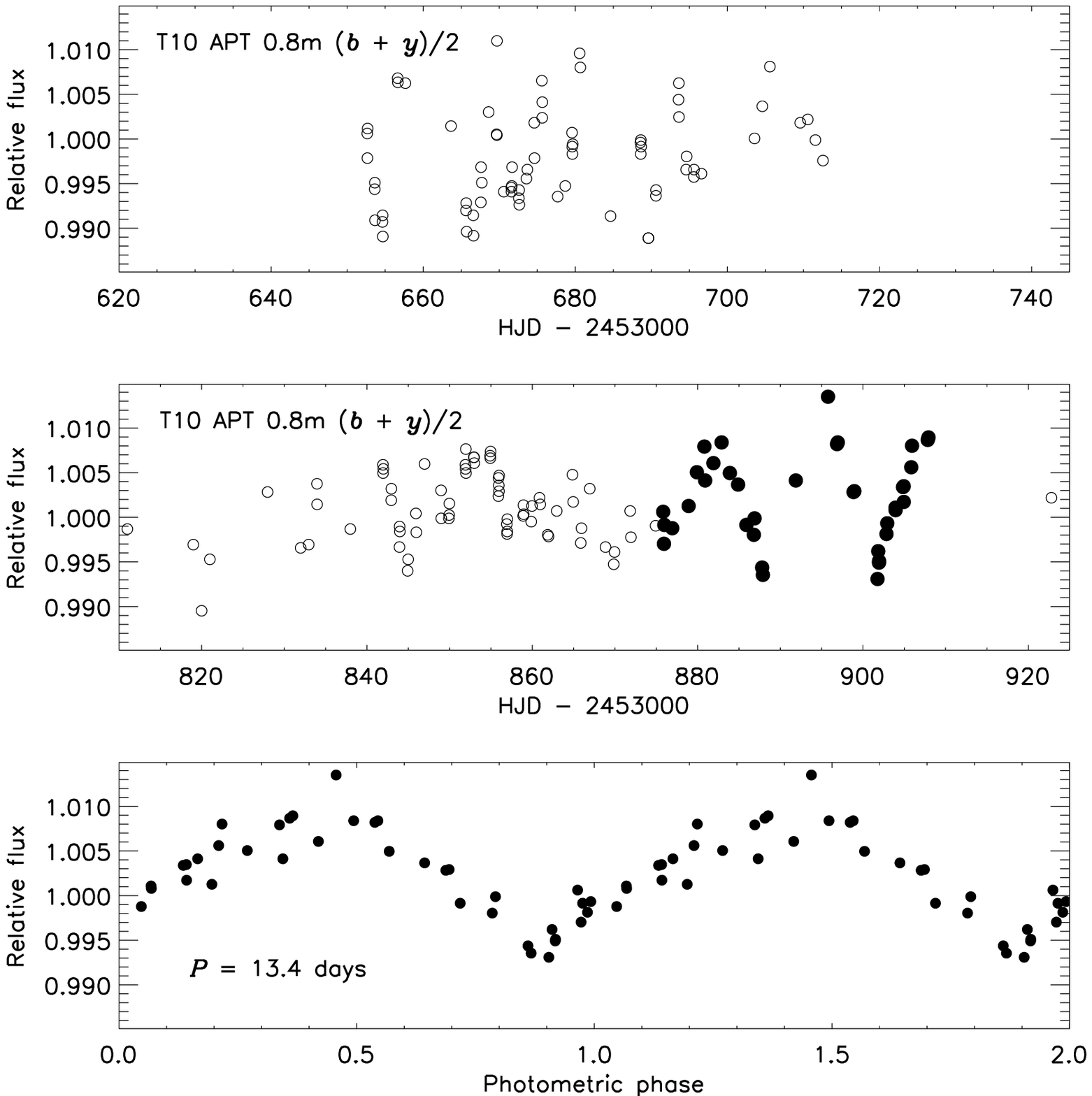}
\caption{
Relative photometry of HD~189733, from the T10 0.8m APT at Fairborn
Observatory. The top two panels show data from the 2005 and 2006
observing seasons, respectively.  The solid symbols show the
portion of data that was used in the periodogram analysis.
This subset of the data is plotted in the bottom panel
as a function of the photometric phase.
\label{fig:3}}
\end{figure}

\subsection{MAGNUM Observatory}

We observed the transit of UT~2006~Aug~21 with the 2m telescope at
the Multi-color Active Galactic NUclei Monitoring (MAGNUM) observatory
on Haleakala, Hawaii (Kobayashi et al. 1998; Yoshii 2002; Yoshii,
Kobayashi, \& Minezaki 2003). This is the same transit that was
observed spectroscopically by Winn et al.~(2006). We used the
multi-color imaging photometer (MIP), which allows for simultaneous
observation with an optical 1024$^2$ SITE CCD and an infrared SBRC
InSb 256$^2$ detector. In this case, we used only the optical
detector, because the infrared detector was saturated even in very
short exposures. We observed in the $V$ band. Because the MIP field of
view is only $1\farcm5 \times 1\farcm5$, and there are no good
comparison stars within this small field, we nodded repeatedly between
the target star and a calibration star, with 1~s exposures. The
calibration star was HD~190449 ($V$ = 8.12, $B-V$ = 0.79, K0). The
average interval between exposures of HD~189733 was 46~s. The median
FWHM of stellar images was $1\farcs2$.

We reduced the images with the standard MIP pipeline described by
Minezaki et al.~(2004). We then performed aperture photometry on
HD~189733 and HD~190449, using an aperture radius of $6\farcs65$ and a
sky annulus ranging in radius from $6\farcs65$ to $9\farcs42$. To
produce a comparison signal, the time series for HD~190449 was
boxcar-smoothed (with a width of 4 points, or 3~minutes) and then
linearly interpolated onto the time stamps of the HD~189733 data.
Then the HD~189733 time series was divided by this comparison signal.
A few extreme outlying points were rejected. To remove residual
systematic errors in the out-of-transit flux determination, we divided
by a linear function of time that was determined as part of the
fitting procedure (see \S~4). The final light curve is shown in the
fifth panel of Fig.~2.

\subsection{Wise Observatory}

We observed the transit of 2006~Sep~5 with the 1m telescope at Wise
Observatory, in Israel. We used a Tektronix 1024$^2$ back-illuminated
CCD detector, giving a pixel scale of $0\farcs7$ and a field of view
of $11\farcm9 \times 11\farcm9$. We observed through a Johnson $I$
filter, the reddest optical band available on this camera. The
exposure time was 10~s and the telescope was defocused in order to
avoid saturation. Autoguiding was used to keep the image registration
constant throughout the night. We also obtained sky flat exposures at
sunset and zero-second (bias) exposures at the beginning and during
the night. We performed the data reduction and photometry using very
similar procedures to those that were used on the FLWO data (\S~2.1).
The bottom panel of Fig.~2 shows the final light curve.

\section{Determination of System Parameters}

To estimate the planetary, stellar, and orbital parameters, and the
times of transit, we fitted a parameterized model to the transit
photometry. The model and the fitting method were similar to those
described in previous TLC papers (see, e.g., Winn et al.~2007),
except that in this case we accounted for correlated noise, as
described below. The model is based on a Keplerian orbit of a star
(with mass $M_\star$ and radius $R_\star$) and a single planet ($M_p$,
$R_p$) about their center of mass. For most of the analysis we assumed
that the orbital eccentricity $e$ is zero, because the expected
timescale of tidal circularization is short in the absence of
excitations from other planets (see, e.g., Rasio et al.~1996, Trilling
et al.~2000, Dobbs-Dixon et al.~2004, Adams \& Laughlin
2006). However, in \S~4.3, we discuss the empirical upper limits on
the orbital eccentricity. The orbit has a period $P$ and an
inclination $i$ relative to the sky plane. We define the coordinate
system such that $0\arcdeg \leq i\leq 90\arcdeg$. It is often useful
to refer to the impact parameter $b \equiv a \cos i / R_\star$ (where
$a$ is the semimajor axis) rather than the inclination.

Because one of our goals was to measure the individual transit times,
we allowed each transit to have an independent value of $T_c$, the
transit midpoint, rather than forcing them to be separated by exact
multiples of the orbital period. Thus, the only effect of $P$ on the
model is to determine the semimajor axis $a$ for a given value of the
total mass. We fixed $P=2.218575$~days, the value determined by Bouchy
et al.~(2005) and H{\'e}brard \& Lecavelier Des Etangs~(2006) from the
detection of transits in the {\it Hipparcos} database. The uncertainty
of $0.000003$~days was negligible for our purposes, although we were
able to use the resulting values of $T_c$ to produce an independent
estimate of the period, as described in \S~5.

Neither $M_\star$ nor $M_p$ can be determined from photometry alone.
As we have done in previous TLC analyses, we fixed $M_\star$ at a
value that is based on an analysis of the stellar spectrum and other
observable properties. We then used the scaling relations $R_p \propto
M_\star^{1/3}$ and $R_\star \propto M_\star^{1/3}$ to estimate the
systematic error associated with the uncertainty in $M_\star$. In this
case we adopted the value $M_\star=0.82\pm 0.03$~$M_\odot$ (Bouchy et
al.~2005). The planetary mass $M_p$ hardly affects the photometric
model at all, but for completeness we used the value
$M_p=1.13$~$M_{\rm Jup}$ (Winn et al.~2006).

To calculate the relative flux as a function of the projected
separation of the planet and the star, we employed the analytic
formulas of Mandel \& Agol~(2002) to compute the integral of the
intensity over the unobscured portion of the stellar disk. We assumed
the limb darkening law to be quadratic,
\begin{equation}
\frac{I_\mu}{I_1} = 1 - u_1(1-\mu) - u_2(1-\mu)^2,
\end{equation}
where $I$ is the intensity, and $\mu$ is the cosine of the angle
between the line of sight and the normal to the stellar surface. We
fixed the limb-darkening coefficients at the values calculated by
Claret~(2000, 2004) for observations of a star with the observed
spectral properties.\footnote{Specifically, we used the tabulated
  values for an ATLAS model with $T_{\rm eff} = 5000$~K, $\log g =
  4.5$~(cgs), log~[M/H]$=0.0$ and $v_t = 2.0$~km~s$^{-1}$. For the $z$
  band, $u_1 = 0.32$ and $u_2 = 0.27$.} We also investigated the
effect of fitting for the limb darkening parameters, as discussed
below. In addition, the light curves exhibited gradients in the
out-of-transit data, probably due to differential extinction between
the target star and the comparison stars, or some other systematic
error. For this reason, each of the 8 data sets was modeled with two
extra parameters: the out-of-transit flux $f_{\rm oot}$ and a time
gradient $\alpha$.

The fitting statistic was
\begin{equation}
\chi^2 =
\sum_{j=1}^{N_f}
\left[
\frac{f_j({\mathrm{obs}}) - f_j({\mathrm{calc}})}{\sigma_j}
\right]^2
,
\label{eq:chi2}
\end{equation}
where $f_j$(obs) is the flux observed at time $j$, $\sigma_j$ controls
the relative weights of the data points, and $f_j$(calc) is the
calculated value.  In order to derive realistic uncertainties on the
parameters, it is important for $\sigma_j$ to include not only
measurement errors but also any unmodeled systematic effects. Of
particular importance is the timescale of the systematic
effects. Correlated noise effectively reduces the number of
independent data points and correspondingly increases the
uncertainties in the model parameters, an issue that Pont et
al.~(2006) and Gillon et al.~(2006) have recently brought to attention
in the context of transit photometry.

Our approach to this problem was as follows. First, for each of the 8
transits, we rescaled the instrumental uncertainties such that
$\chi^2/N_{\rm dof} = 1$ for the best-fitting model. The resulting
uncertainties are those that are given in Table~2. Second, we followed
the procedure of Gillion et al.~(2006) to decompose the observed noise
into ``white noise'' (that which averages down as $1/\sqrt{N}$, where
$N$ is the number of data points) and ``red noise'' (that which does
not average down over some specified time interval). Specifically, we
calculated the standard deviation of the residuals ($\sigma$) and the
standard deviation of the time-averaged residuals ($\sigma_N$). The
averaging time was 1~hr (a timescale comparable to the transit event),
corresponding to a number $N$ of data points that depended upon the
cadence of observations.  Then we solved for the white noise
$\sigma_w$ and red noise $\sigma_r$ from the system of equations
\begin{eqnarray}
\sigma_1^2 & = & \sigma_w^2 + \sigma_r^2, \\
\sigma_N^2 & = & \frac{\sigma_w^2}{N} + \sigma_r^2.
\end{eqnarray}
Finally, to account approximately for the effective reduction in the
number of independent data points, we rescaled the $\sigma_j$ in
Eq.~(\ref{eq:chi2}) by the factor $\sigma_r/(\sigma_w/\sqrt{N})$.

The results for $\sigma_w$, $\sigma_r$, and the rescaling factor for
each data set are given in Table 1. These results are not very
sensitive to the choice of averaging time. Any choice between 15~min
and a few hours gave similar results. Among the data sets are wide
disparities in the degree of red noise, ranging over a factor of
10. By far the best data, in the sense of the smallest noise
correlations, are from the FLWO 1.2m telescope and Keplercam
(\S~2.1). For this reason, we decided to estimate the system
parameters using {\it only} the FLWO data, and use the other data sets
only to determine transit times and as a consistency check on the FLWO
results.  (Had we included the other data sets, their statistical
weight would anyways have been much smaller.)

In short, we used the FLWO data to solve for the two bodies' radii
($R_\star$ and $R_p$); the orbital inclination ($i$); and the
mid-transit time ($T_c$), the out-of-transit flux ($f_{\rm oot}$), and
a time gradient ($\alpha$) for each of the 2 transits.  We then fixed
$R_\star$, $R_p$, and $i$ at the best-fitting values, and fitted each
of the remaining 6 data sets to find $T_c$, $f_{\rm oot}$, and
$\alpha$.

To solve for the model parameters and their uncertainties, we used a
Markov Chain Monte Carlo algorithm (see, e.g., Tegmark 2004). Our jump
function was the addition of a Gaussian random number to each
parameter value. We set the perturbation sizes such that $\sim$20\% of
jumps are executed. We created 10 independent chains, each with
500,000 points, starting from random initial positions, and discarded
the first 20\% of the points in each chain. The Gelman \& Rubin~(1992)
$R$ statistic was within 0.2\% of unity for each parameter, a sign of
good mixing and convergence. We merged the chains and took the median
value of each parameter to be our best estimate, and the standard
deviation as the 1~$\sigma$ uncertainty.

\section{Results}

The results are given in Table~3. Along with the results for the model
parameters, we have provided results for some useful derived
quantities such as the impact parameter $b$, the radius ratio
$(R_p/R_\star)$, and the fraction $(R_p/a)^2$ (which gives the
fraction of starlight reflected by the planet, for an albedo of
unity). We also report the calculated values of the full transit
duration (the time between first and fourth contact, $t_{\rm IV} -
t_{\rm I}$), and the partial transit duration (the time between first
and second contact, or between third and fourth
contact).\footnote{Although the partial transit duration is listed as
  $t_{\rm II} - t_{\rm I}$ in Table~3, all of the results in Table~1
  are based on the entire light curves, including both ingress and
  egress data. Our model assumes $t_{\rm II} - t_{\rm I} = t_{\rm IV}
  - t_{\rm III}$.}

\subsection{Stellar and Planetary Radii}

The result for the stellar radius is $R_\star = 0.753 \pm
0.025~R_\odot$. The uncertainty is dominated by the statistical error
of 0.023~$R_\odot$ (3.1\%). The covariance with the uncertainty in the
stellar mass produces an additional error of 0.009~$R_\odot$ (1.2\%),
which we have added in quadrature to the statistical error to arrive
at the net uncertainty of 0.025~$R_\odot$.  We find the planetary
radius to be $R_p = 1.156\pm 0.046~R_{\rm Jup}$, where the uncertainty
is again dominated by the statistical error of $0.044~R_{\rm Jup}$.

To test the robustness of these results, we performed some additional
fits. We gauged the importance of the choice of limb-darkening law by
re-fitting the data under different assumptions.  When we allowed the
limb-darkening coefficients $u_1$ and $u_2$ to be free parameters
rather than holding them fixed, we found $R_\star = 0.755~R_\odot$ and
$R_p = 1.163~R_{\rm Jup}$, well within the 1~$\sigma$ error of our
original analysis.  (The optimized limb-darkening coefficients for the
$z$ band were $u_1=0.35$ and $u_2=0.22$, as compared to the
theoretical values of $u_1=0.32$ and $u_2=0.27$.) Likewise, the
results changed by only 0.25~$\sigma$ when we used a linear
limb-darkening law ($u_2=0$), regardless of whether the linear
limb-darkening coefficient was fixed or taken to be a free parameter.
As another check, we fitted each of the 8 transit data sets
separately, solving for the parameters $\{R_\star, R_p, i, T_c,
\alpha, f_{\rm oot}\}$ in each case. Although the statistical power of
the FLWO photometry was the greatest, as noted previously, all 8 sets
of results agreed within their calculated error bars. In particular,
the unweighted ``ensemble averages'' of the results from the 6
non-FLWO data sets were $R_\star = 0.760~R_\odot$ and $R_p =
1.148~R_{\rm Jup}$, again in agreement with our original analysis.

How do these results compare to the previous analyses of HD~189733 by
Bouchy et al.~(2005) and Bakos et al.~(2006a)?  An important
difference is that those authors decided not to determine $R_\star$
from the transit photometry.\footnote{While Bakos et al.~(2006a) did
fit for the stellar radius, finding $R_\star = 0.68\pm 0.02~R_\odot$,
they did not trust the result. They quite reasonably suspected that
the true error bar was significantly larger than their calculations
indicated, because of correlated noise in the data.}  Rather, those
authors used estimates of $R_\star$ based on an analysis of other
observable properties of the star. Bouchy et al.~(2005) used
measurements of the star's parallax, effective temperature, surface
gravity, and metallicity, in comparison with the outputs of stellar
evolution models, and concluded $R_\star = 0.76\pm
0.01~R_\odot$. Likewise, Bakos et al.~(2006a) investigated four
different ways of determining the stellar radius, based on broad band
colors, spectral properties, and model isochrones, and found a stellar
radius in the range 0.74--0.79~$R_\odot$. Ultimately, Bakos et
al.~(2006a) adopted the value $0.758\pm 0.016~R_\odot$ based on a
calibration of 2MASS photometry by Masana et al.~(2006). Our
determination of $R_\star$ based on the transit light curve is in
agreement with those independent determinations; the mutual agreement
constitutes an important consistency check on the data and our
analysis.

Our value for the planetary radius agrees with the value $R_p =
1.154\pm 0.033~R_{\rm Jup}$ found by Bakos et al.~(2006a). While the
Bakos et al.~(2006a) result would appear to be more precise, the
comparison is somewhat misleading because Bakos et al.~(2006a) were
not simultaneously fitting for $R_\star$, as noted above. If we follow
their procedure of fixing $R_\star = 0.758\pm 0.016~R_\odot$, then our
result for the planetary radius becomes more precise: $R_p = 1.164 \pm
0.028~R_{\rm Jup}$.  Both our result and that of Bakos et al.~(2006a)
disagree with that of Bouchy et al.~(2005), who found $1.26\pm
0.03~R_{\rm Jup}$ based on a $B$-band light curve from the 1.2m
telescope at the Observatoire de Haute-Provence (OHP).  This
discrepancy was traced by Bakos et al.~(2006a) to systematic errors in
the OHP photometry. When the $B$ band light curve was recalculated
using a greater number of comparison stars, the transit depth
decreased by 20\% and the inferred planetary radius shrank
accordingly.

How does the planetary radius compare to theoretical expectations,
given its mass ($M_p = 1.13 \pm 0.03~M_{\rm Jup}$; Winn et al.~2006)
and its proximity to its parent star ($a=0.035$~AU)?  Fortney, Marley,
\& Barnes (2006) have recently provided a wide range of theoretical
predictions for exoplanet radii.  Their calculations are for a
solar-mass star, but at their suggestion we can rescale the semimajor
axis to compensate for the lower luminosity of HD~189733.  Assuming
$(L_\star/L_\odot) = (M_\star/M_\odot)^{3.5}$, then if HD~189733 were
orbiting the Sun at $a=0.05$ it would receive roughly the same
incident flux as it does in its actual orbit. The resulting prediction
from Fig.~6 of Fortney et al.~(2006) is a planetary radius between
$\approx$1.05--1.12~$R_{\rm Jup}$, assuming an age of $\sim$4.5~Gyr
and depending on whether or not the planet has a massive
($25~M_\oplus$) core. The core-free prediction for the radius is
larger and in agreement with the observed value. However, it seems
premature to claim that a massive core is disfavored, given the
uncertainties that enter into the calculations and the uncertainty in
the age of the system. It does seem safe to say that the radius of
HD~189733 does not present a severe theoretical problem, unlike the
cases of the apparently ``bloated'' planets HD~209458b, HAT-P-1b, and
WASP-1b (for recent results on those systems, see Knutson et al.~2006,
Bakos et al.~2006b, Collier-Cameron et al.~2006, and Charbonneau et
al.~2006b).

The mean density of HD~189733b is $\rho_p = 0.91 \pm
0.06$~g~cm$^{-3}$, which is between the densities of Saturn
(0.6~g~cm$^{-3}$) and Jupiter (1.2~g~cm$^{-3}$). The surface gravity
of HD~189733b is $g = 21\pm 1$~m~s$^{-2}$, which is also intermediate
between Saturn (10~m~s$^{-2}$) and Jupiter (25~m~s$^{-2}$). We note
that whenever $M_p$ is measured via the spectroscopic orbit of the
star, and $R_p$ is measured via transit photometry (as is the case
here), then the derived value of $g$ is immune to systematic errors in
the parameters of the parent star. This is because the fitting
degeneracies are $M_p \propto M_\star^{2/3}$ and $R_p \propto
M_\star^{1/3}$, and hence $g \equiv GM_p/R_p^2$ is independent of
$M_\star$. Here this fact is of limited interest, because the error in
$R_p$ is dominated by statistical error, but it may be of importance
in future transit studies.\footnote{We thank S.~Gaudi for helping us
  to appreciate this point.}

\subsection{Determination of the Transit Ephemerides}

Table~4 gives the 8 transit times measured from our data. We have used
these times, along with transit times previously measured by Bakos et
al.~(2006), to calculate a photometric ephemeris for this system.
Although Bakos et.~(2006) reported 15 measured times, we used only 4
of those data points in our analyis. We did not include the 10 times
that were based on only partial observations of the transit.  Full
transits are greatly preferable, in order to correct (or at least
assess) systematic errors using the pre-ingress and post-egress data.
In addition, we did not include the $T_c$ measurement based on the OHP
$B$-band light curve of Bouchy et al.~(2005) because of the systematic
errors noted previously.  What remained were 4 data points
representing 4 independent measurements of the same event.  In
combination with our 8 data points, we fitted a linear function of
transit epoch $E$,
\begin{equation}
T_c(E) = T_c(0) + E P,
\label{eq:ephemeris}
\end{equation}
finding $T_c(0) = 2453988.80336(23)$~[HJD] and $P =
2.2185733(19)$~days, where the numbers in parentheses indicate the
$1~\sigma$ uncertainty in the final two digits. The fit had
$\chi^2/N_{\rm dof} = 1.08$ and $N_{\rm dof} = 11$.  We chose $E=0$ to
correspond to the most precisely known transit time. Our derived
period agrees almost exactly with the value $2.2185730(20)$~days
determined by Bakos et al.~(2006a), and it is also in agreement with
the {\it Hipparcos}-based values of $2.2185750(30)$~days (Bouchy et
al.~2005) and $2.218574 ^{+0.000006}_{-0.000010})$~days (H\'{e}brard
\& Lecavelier des Estangs 2006).

Figure~4 is the O$-$C (observed minus calculated) diagram for the
transit times, according to this new ephemeris.  The filled symbols
represent data points used in the fit.  There is not yet any pattern
in the residuals that would indicate the effect of a perturbing body
in the system.  The unfilled square shows the OHP $B$-band
measurement, which is indeed an outlier.  The unfilled circle is
explained in the next section.

\begin{figure}[p]
\epsscale{1.0}
\plotone{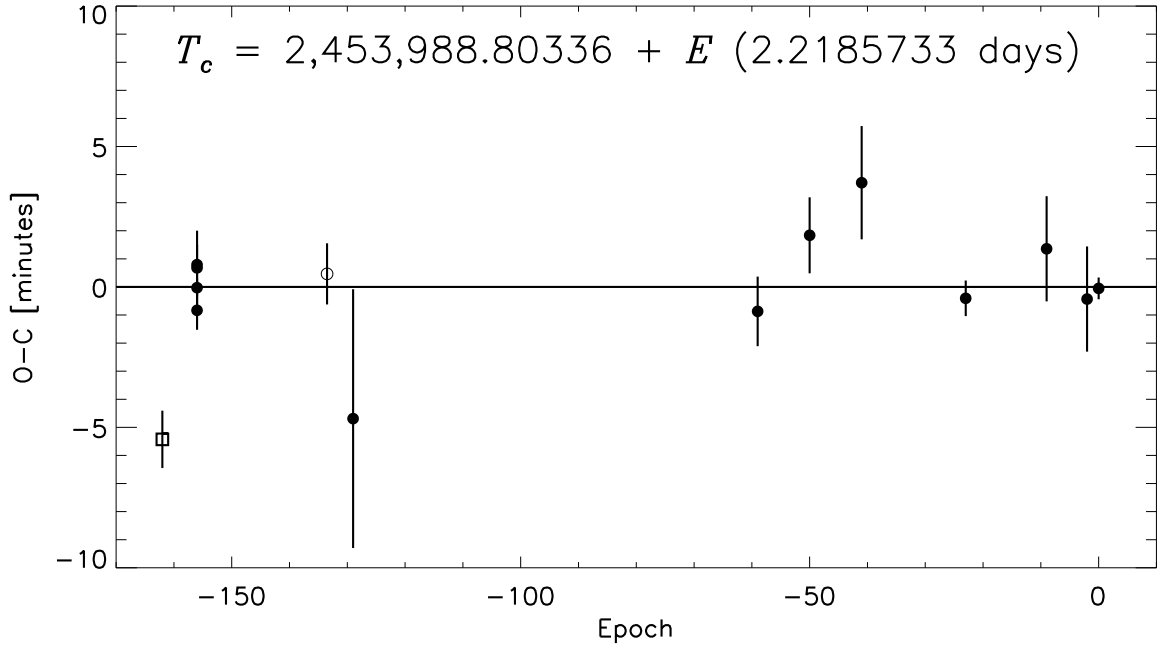}
\vskip 0.2in
\caption{ Transit and secondary-eclipse timing residuals for
HD~189733.  The calculated times, using the ephemeris of
Eq.~(\ref{eq:ephemeris}), have been subtracted from the observed
times.  The filled symbols represent data points used in the fit. The
unfilled square is based on $B$-band data by Bouchy et al.~(2005), as
re-analyzed by Bakos et al.~(2006a). The unfilled circle is the
secondary-eclipse time measured by Deming et al.~(2006).
\label{fig:4}}
\end{figure}

\subsection{Limits on the Orbital Eccentricity}

As mentioned previously, one would expect the orbit of a hot Jupiter
such as HD~189733b to be very nearly circular, due to tidal effects.
Previous results and our results have all shown that a circular orbit
does indeed provide a satisfactory description of the available data.
However, it is still interesting to make an empirical determination of
the eccentricity, both in the spirit of ``what one can measure, one
should measure'' and also because any additional bodies in the system
could excite the orbital eccentricity. We used a two-step procedure to
determine the orbital eccentricity.

First, we used our revised ephemeris to interpret the
secondary-eclipse timing of Deming et al.~(2006).  For a nonzero (but
small) orbital eccentricity, the time difference between the midpoint
of secondary eclipse, $T_{\rm sec}$, and the time of transit, $T_{\rm
  tra}$, may differ from half of the orbital period:
\begin{equation}
T_{\rm sec} - T_{\rm tra} \approx \frac{P}{2} \left( 1 + \frac{4}{\pi}~e\cos\omega \right),
\label{eq:ecosw}
\end{equation}
where $\omega$ is the argument of pericenter. Deming et al.~(2006)
measured the midpoint of a secondary eclipse at HJD~$2453692.62416\pm
0.00067$, which is consistent with the $e=0$ prediction of our
ephemeris. The unfilled circle in Fig.~4 represents the secondary
eclipse measurement. The timing offset of Eq.~(\ref{eq:ecosw}) is
$0.46\pm 1.1$~minutes, corresponding to $e\cos\omega = 0.00023\pm
0.00054$.

Second, to determine the other component $e\sin\omega$ of the
eccentricity vector, we used the radial velocities presented by Winn
et al.~(2006).  We used only those 60 velocities measured outside of
transits (i.e., not affected by the Rossiter-McLaughlin effect). We
performed an MCMC analysis to solve for the Keplerian orbital
parameters, as well as a possible long-term velocity gradient, using
the same treatment of the measurement errors that Winn et al.~(2006)
applied to the entire data set. We allowed both $e$ and $\omega$ to be
free parameters, but with an a priori constraint on $e\cos\omega$ to
enforce compliance with the secondary-eclipse measurement.  The result
for $e\sin\omega$ was $-0.007\pm 0.011$.

Hence, both components of the eccentricity vector are consistent with
zero, and $e\cos\omega$ is about 20 times more tightly bounded than
$e\sin\omega$.  Values of $e$ as large as $\approx$0.02 are allowed,
but only for $\omega$ very close to $\pm 90\arcdeg$.

\subsection{Three-Dimensional Spin-Orbit Alignment}

Thanks to the APT data (\S~2.2), HD~189733 is the first star with a
measured rotation period that also has a transiting planet. Together
with the transit photometry and the observation of the
Rossiter-McLaughlin effect, this allows for the determination of the
angle between the stellar rotation axis and the planetary orbit
normal, as anticipated by Queloz et al.~(2000). This angle is worth
measuring because any significant misalignment may be an indication of
perturbative effects during planetary migration, among other reasons
(as explained in more detail by Ohta et al.~2005, Winn et al.~2005 and
Gaudi \& Winn 2007).

The true (three-dimensional) angle $\psi$ between the stellar spin
axis and the orbital axis is given by the formula
\begin{equation}
\cos\psi = \cos i_\star \cos i + \sin i_\star \sin i \cos\lambda,
\label{eq:angles}
\end{equation}
where $i$ is the orbital inclination, $i_\star$ is the inclination of
the stellar rotation axis, and $\lambda$ is the angle between the sky
projections of the two axes. (For a diagram of the coordinate system,
see Fig.~3 of Ohta et al.~2005.) The transit photometry determines $i$
with excellent accuracy. Observations of the Rossiter-McLaughlin
effect have been used to determine $\lambda$, but they cannot be used
to determine $i_\star$ independently. Rather, they are sensitive to
$v\sin i_\star$, the projected rotation rate of the star. Given $v\sin
i_\star$ from the Rossiter-McLaughlin observations\footnote{Of course
  it is also possible to use a more traditional measurement of $v\sin
  i_\star$, from an analysis of the width of photospheric absorption
  lines.}, along with $i$ and $R_\star$ from the transit photometry,
and the stellar rotation period $P_{\rm rot}$, one can determine $\sin
i_\star$ via the formula
\begin{equation}
\sin i_\star = v\sin i_\star \left( \frac{P_{\rm rot}}{2\pi R_\star} \right).
\end{equation}
Hence, all of the angles in Eq.~(\ref{eq:angles}) are known. Using the
values $\lambda = -1\fdg4\pm 1\fdg1$, and $v\sin i_\star = 2.97\pm
0.22$~km~s$^{-1}$ from Winn et al.~(2006), we found $\sin i_\star =
1.04 \pm 0.09$. By rejecting values of $\sin i_\star > 1$ as
unphysical, and propagating the errors through Eq.~(\ref{eq:angles}),
we determined an upper bound on the (mis)alignment angle $\psi$ of
$27\arcdeg$ with 95\% confidence.  Essentially the same result can be
obtained from the approximation $\cos\psi \approx \sin i_\star$, which
is valid because $i\approx 90\arcdeg$ and $\lambda\approx 0\arcdeg$
are tightly constrained.

This is the first exoplanetary system for which it has been possible
to measure $\psi$. The result is consistent with zero, but it is not
as precise as the result for the projected angle $\lambda$.  How could
the measurement of $\psi$ be improved?  We have already mentioned some
caveats relating to the measurement of the rotation period\footnote{We
  note that if the logic in this section is reversed, and one is
  willing to {\it assume} perfect spin-orbit alignment, then the
  calculated rotation period is $P_{\rm rot} = 12.8\pm 1.0$~days.}
(\S~2.2), but the current uncertainty in $\psi$ is dominated by the
error in $v\sin i_\star$ which is itself dominated by systematic
errors arising from the interpretation of the transit spectra (see
Winn et al.~2006).  Specifically, the systematic error arose from the
``calibration'' the Rossiter-McLaughlin effect using simulated
spectra, which was needed because the transit spectra were observed
through an I$_2$ cell and analyzed with an algorithm that is nominally
designed for measuring Doppler shifts rather than spectral
distortions.  Further improvement might be achieved through a more
sophisticated set of simulations or perhaps by re-observing the
Rossiter-McLaughlin effect without the I$_2$ cell (i.e., using a
different technique to account for instrumental variations).

\section{Summary}

We have presented photometry of 8 complete transits of the exoplanet
HD~189733b, and modeled the light curves in order to determine the
radii of the star and the planet. Our results are consistent with
previous results and with theoretical expectations for close-in Jovian
planets. Stringent limits on the orbital eccentricity follow from the
measured transit times, in conjunction with a previous detection of
the secondary eclipse and with the spectroscopic orbit.  We have also
presented nightly out-of-transit photometry spanning 2~yr that has
revealed the stellar rotation period. We have used this information,
along with a previous analysis of the spectroscopic transit, to place
an upper bound on the true angle between the stellar rotation axis and
the orbital axis. With these developments, HD~189733b has become one
of the most thoroughly characterized planets outside of the Solar
system.

\acknowledgments We thank F.~Pont and M.~Gillon for helpful
discussions about correlated noise. We are grateful to G.~Marcy,
P.~Butler, S.~Vogt, and E.~Turner for their help with the Doppler
analysis and for encouragement. A.R.~thanks the MIT UROP office for
research funding.

\begin{deluxetable}{lcccccc}
\tabletypesize{\small}
\tablecaption{Characteristics of Transit Data\label{tbl:data}}
\tablewidth{0pt}

\tablehead{
\colhead{Date} & \colhead{Telescope} & \colhead{Filter} & \colhead{Cadence} &
\colhead{White Noise} & \colhead{Red Noise} & \colhead{Reweighting factor} \\
\colhead{(UT)} & \colhead{} & \colhead{} & \colhead{(min)} &
\colhead{$\sigma_w$} & \colhead{$\sigma_r$} & \colhead{$\sigma_r/(\sigma_w/\sqrt{N_{\rm tr}})$}
}

\startdata
2005~Nov~28 & T10 APT 0.8m & $(b+y)/2$ & 1.44 & 0.0045 & 0.00250 & 4.2 \\
2006~May~2  & T10 APT 0.8m & $(b+y)/2$ & 1.44 & 0.0024 & 0.00080 & 2.8 \\
2006~May~22 & T10 APT 0.8m & $(b+y)/2$ & 1.44 & 0.0024 & 0.00090 & 3.0 \\
2006~Jun~11 & T10 APT 0.8m & $(b+y)/2$ & 1.30 & 0.0019 & 0.00140 & 5.3 \\
2006~Jul~21 & FLWO 1.2m & $z$       & 0.23 & 0.0023 & 0.00035 & 3.3 \\
2006~Aug~21 & MAGNUM 2m & $V$       & 1.38 & 0.0029 & 0.00100 & 2.9 \\
2006~Sep~5  & Wise 1m   & $I$       & 0.73 & 0.0029 & 0.00120 & 4.6 \\
2006~Sep~10 & FLWO 1.2m & $z$       & 0.23 & 0.0022 & 0.00020 & 2.0
\enddata

\end{deluxetable}

\begin{deluxetable}{lcccc}
\tabletypesize{\normalsize}
\tablecaption{Photometry of HD~189733\label{tbl:photometry}}
\tablewidth{0pt}

\tablehead{
\colhead{HJD} & \colhead{Telescope} & \colhead{Filter} &
\colhead{Relative flux} & \colhead{Uncertainty}
}

\startdata
      FLWO 1.2m &        $z$ &   2453937.71893 &          1.0037 &          0.0027 \\
      FLWO 1.2m &        $z$ &   2453937.71909 &          0.9963 &          0.0027 \\
      FLWO 1.2m &        $z$ &   2453937.71925 &          1.0007 &          0.0027
\enddata 

\tablecomments{The time stamps represent the Heliocentric Julian Date
  at the time of mid-exposure. We intend for this Table to appear in
  entirety in the electronic version of the journal. A portion is
  shown here to illustrate its format. The data are also available
  from the authors upon request.}

\end{deluxetable}

\begin{deluxetable}{lcc}
\tabletypesize{\normalsize}
\tablecaption{System Parameters of HD~189733\label{tbl:params}}
\tablewidth{0pt}

\tablehead{
\colhead{Parameter} & \colhead{Value} & \colhead{Uncertainty}
}

\startdata
            $(R_\star/R_\odot)(M_\star/0.82~M_\odot)^{-1/3}$ & $          0.753$ & $          0.023 $ \\
            $(R_p/R_{\rm Jup})(M_\star/0.82~M_\odot)^{-1/3}$ & $          1.156$ & $          0.044 $ \\
                                           $R_\star/R_\odot$ & $          0.753$ & $          0.025 $ \\
                                           $R_p/R_{\rm Jup}$ & $          1.156$ & $          0.046 $ \\
                                             $R_p / R_\star$ & $         0.1575$ & $         0.0017 $ \\
                                                 $(R_p/a)^2$ & $       0.000313$ & $       0.000025 $ \\
                                                 $R_\star/a$ & $         0.1124$ & $         0.0034 $ \\
                                                   $i$~[deg] & $          85.76$ & $           0.29 $ \\
                                                         $b$ & $          0.658$ & $          0.027 $ \\
                               $t_{\rm IV} - t_{\rm I}$~[hr] & $          1.827$ & $          0.029 $ \\
                              $t_{\rm II} - t_{\rm I}$~[min] & $           24.6$ & $            1.9 $ \\
\enddata

\end{deluxetable}

\begin{deluxetable}{lcc}
\tabletypesize{\normalsize}
\tablecaption{Mid-transit times of HD~189733b\label{tbl:times}}
\tablewidth{0pt}

\tablehead{
\colhead{Epoch} & \colhead{Mid-transit time} & \colhead{Uncertainty} \\
\colhead{$E$}   & \colhead{[HJD]}            & \colhead{[days]}      
}

\startdata
$-129$ & $  2453702.60416$ & $        0.0032  $ \\
$-59$  & $  2453857.90694$ & $        0.00086 $ \\
$-50$  & $  2453877.87598$ & $        0.00094 $ \\
$-41$  & $  2453897.84444$ & $        0.0014  $ \\
$-23$  & $  2453937.77590$ & $        0.00044 $ \\
$-9$   & $  2453968.83715$ & $        0.0013  $ \\
$-2$   & $  2453984.36592$ & $        0.0013  $ \\
$0$    & $  2453988.80331$ & $        0.00027 $ \\
\enddata

\end{deluxetable}

\end{document}